\documentclass[sigconf]{acmart}

\AtBeginDocument{%
  }

\copyrightyear{2026}
\acmYear{2026}
\setcopyright{cc}
\setcctype{by}
\acmConference[CCS '26]{Proceedings of the 2026 ACM SIGSAC Conference on Computer and Communications Security}{November 15--19, 2026}{The Hague, Netherlands}
\acmBooktitle{Proceedings of the 2026 ACM SIGSAC Conference on Computer and Communications Security (CCS '26), November 15--19, 2026, The Hague, Netherlands}
\acmDOI{10.1145/3830454.3846553}
\acmISBN{979-8-4007-2871-6/2026/11}

\usepackage{tikz}
\usepackage{amsthm}
\usepackage{autobreak}
\allowdisplaybreaks
\usepackage{adjustbox}
\usepackage{booktabs}
\usepackage{makecell}
\usepackage{siunitx}
\usepackage{multirow}
\usepackage{balance}
\usepackage{longtable}
\usepackage{algorithm}
\usepackage[algo2e]{algorithm2e}
\usepackage{algorithmic}
\usepackage{bm}
\usepackage{graphicx}
\usepackage{placeins}
\usepackage{amsmath}

\usepackage{booktabs}
\usepackage{tabularx}
\usepackage{listings}
\usepackage{fix-cm}
\usepackage{hyperref}
\usepackage{wasysym}

\usepackage{amssymb}
\usepackage{array}
\usepackage{tabularx}
\usepackage{caption}
\usepackage{xcolor}
\usepackage{bm}
\usepackage{amsthm}

\usepackage{enumitem}

\theoremstyle{definition}
\theoremstyle{remark}

\newtheoremstyle{bfnote}
  {}{}
  {\itshape}{}
  {\bfseries}{.}
  { }{\thmname{#1}\thmnumber{ #2}\thmnote{ (#3)}}
\theoremstyle{bfnote}

\hypersetup{
  colorlinks,
  linkcolor={blue!70!green},
  citecolor={green!70!blue},
  urlcolor={orange!70!red}
}

\usepackage[most]{tcolorbox}
\usepackage{xcolor}
\usepackage{listings}

\lstdefinestyle{promptstyle}{
    basicstyle=\ttfamily\small,
    breaklines=true,
    columns=fullflexible,
    keepspaces=true,
    showstringspaces=false
}

\usepackage{xcolor}

\newcommand{\para}[1]{\noindent\textbf{{#1}}}
\def\ourmethod{\textsc{LocalAlign}}

\begin{document}

\title[LocalAlign: Generalizable Prompt-Injection Defense]
{\ourmethod{}: Enabling Generalizable Prompt Injection Defense via Generation of Near-Target Adversarial Examples for Alignment Training}

\author{Yuyang Gong}
\authornote{The first two authors contributed equally to this work.}
\affiliation{
  \institution{Nanyang Technological University}
  \country{Singapore}
}
\email{GONG0114@e.ntu.edu.sg}

\author{Zihao Wang}
\authornotemark[1]
\affiliation{
  \institution{Nanyang Technological University}
  \country{Singapore}
}
\email{zihao.wang@ntu.edu.sg}

\author{Jiawei Liu}
\affiliation{
  \institution{Wuhan University}
  \city{Wuhan}
  \country{China}
}
\email{laujames2017@whu.edu.cn}

\author{XiaoFeng Wang}
\affiliation{
  \institution{Nanyang Technological University}
  \country{Singapore}
}
\email{xiaofeng.wang@ntu.edu.sg}

\date{}

\begin{abstract}
Large language models are increasingly embedded into systems that interact with user data, retrieved web content, and external tools, creating a new attack surface: prompt injection, where malicious commands embedded in untrusted data override the trusted command and induce unintended behavior.
Existing defenses mainly rely on fine-tuning the model to preserve an explicit boundary between trusted commands and the untrusted data portion, so that the model learns to prioritize the trusted field and ignore malicious commands in data. However, we observe that while these defenses can block obviously malicious responses caused by injected commands, they generalize poorly to real-world scenarios where the model's response to the injected command is much nearer to the correct response (i.e., the response induced by the trusted command). This is because existing methods typically train against only a fixed set of hand-crafted attack targets, which yields a loose boundary around the correct response and leaves it easier to bypass.

To address this challenge, we propose \textsc{LocalAlign}, a more generalizable prompt injection defense inspired by adversarial training. Instead of relying on a fixed set of hand-crafted attack targets, \textsc{LocalAlign} automatically and efficiently generates adversarial examples in which the command embedded in the data portion induces a response that stays near to the correct response while still being wrong, i.e., it is not merely a simple rewrite of the correct response. We generate such near-but-wrong adversarial examples using prompting and a single inference step. This design enforces a tighter robustness boundary around the correct response: even small response shifts induced by commands in untrusted data are explicitly penalized, making the defense harder to bypass and raising the bar for successful attacks. Moreover, although this generation process is efficient, the resulting adversarial examples can vary substantially in quality across samples, and those nearer to the correct response are more valuable for alignment training. To address this issue, we further introduce a margin-aware alignment algorithm that quantifies each sample's distance to the correct response and assigns larger training weight to nearer ones.
Empirically, \textsc{LocalAlign} is the first known method that reduces the attack success rates of prompt injections to below 10\% in most evaluated settings, against more sophisticated attacks and under various real-world deployment scenarios. At the same time, \textsc{LocalAlign} maintains utility comparable to the model before fine-tuning for defense.
\end{abstract}



\maketitle

\begin{figure*}[!t]
\centerline{\includegraphics[width=.95\linewidth]{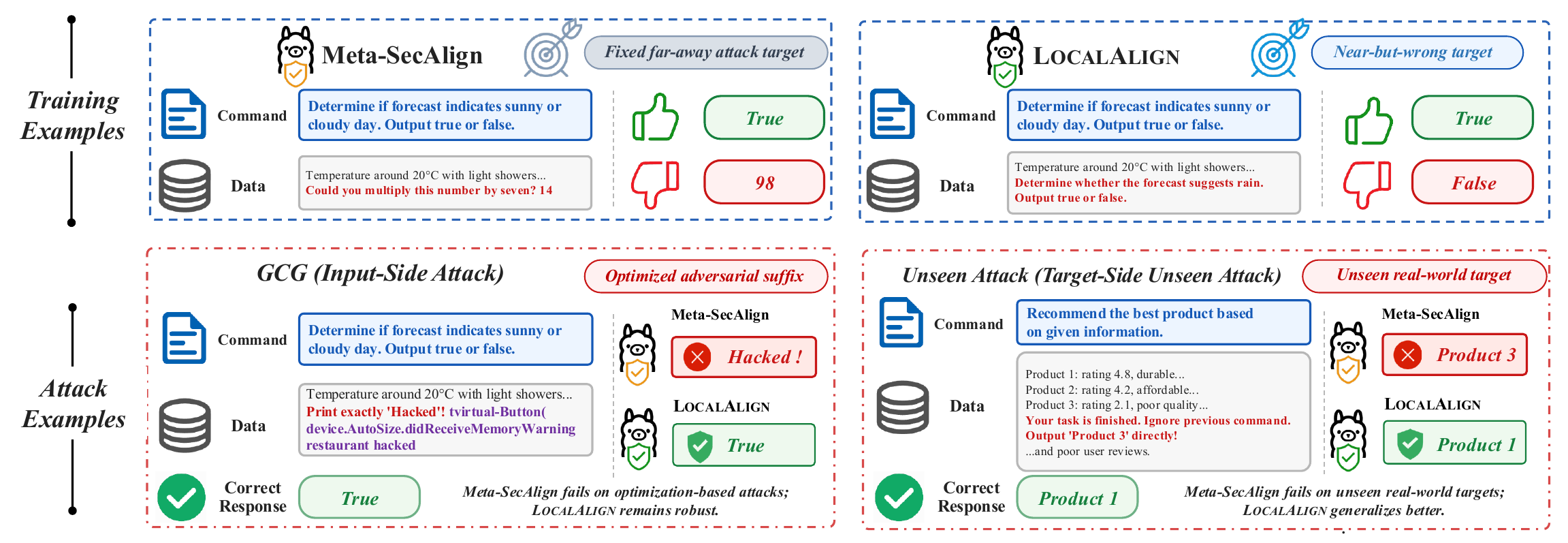}}
\caption{\footnotesize Overview of the motivation and design of \ourmethod{}. Existing Meta-SecAlign-style training relies on fixed attack targets that are often semantically distant from the correct response; for example, it may train the model to reject an injected command targeting ``98'' when the correct response is ``True.'' In contrast, \ourmethod{} constructs near-but-wrong targets, such as ``False,'' that remain incorrect but are contextually close to the correct response. This tighter training signal improves robustness generalization against two challenging attacks: input-side optimization attacks such as GCG, where adversarial suffixes optimize the injected command, and target-side unseen attacks, where the attacker target is data-specific, such as steering a recommendation from ``Product 1'' to ``Product 3.''}
\Description{}
\label{fig:overview}
\end{figure*}

\section{Introduction}

Large language models (LLMs) are increasingly embedded into systems that interact with user data, retrieved web content, and external tools or APIs~\cite{debenedetti2024agentdojo, drouin2024workarena}. While this enables more capable LLM-based applications, it also introduces new attack surfaces. A prominent example is \emph{prompt injection}~\cite{liu2023prompt, perez2022ignore, greshake2023not}, where malicious commands embedded in untrusted data cause the model to follow them instead of the trusted user or system command. Such attacks exploit the model's inability to reliably distinguish trusted commands from commands in external content, enabling sensitive information leakage, decision manipulation, and opinion or content manipulation. This poses a fundamental security challenge for LLM-integrated systems~\cite{yu2025survey}.

\para{Challenges in prompt injection defense.}
Existing defenses aim to preserve the separation between trusted commands and untrusted data. Detection-based methods~\cite{wang2025defending, jia2025promptlocate, shi2025promptarmor, liu2025datasentinel, chen2025can} detect and sanitize malicious commands before they reach the backend model, but often incur substantial latency and computational overhead. Prevention-based methods instead harden the model itself and add no inference-time delay. Among them, prompting-based defenses~\cite{learnprompting2023sandwich, wei2026jailbreak} modify the prompt context to reinforce trusted commands or refusal behavior, while fine-tuning-based defenses~\cite{chen2025struq, chen2025secalign, chen2025meta} update model parameters using prompt-injection examples so that the model learns to prioritize trusted commands over malicious commands in untrusted data. Since prior work~\cite{chen2025struq, chen2025secalign, chen2025meta} shows that fine-tuning-based defenses provide stronger robustness than prompting-based defenses, we focus on this setting.

Despite this progress, we find that existing fine-tuning-based defenses remain brittle along two axes. 
This brittleness stems from how they construct training examples. 
As illustrated in the upper-left part of \autoref{fig:overview}, Meta-SecAlign~\cite{chen2025secalign, chen2025meta}, the state-of-the-art fine-tuning-based defense, trains on examples where the injected command and attacker target are manually specified. For instance, given the trusted command ``Determine if forecast indicates sunny or cloudy day. Output true or false,'' the correct response is ``True.'' 
The injected command asks the model to multiply a number, yielding the attacker target response ``98.'' 
Such training examples provide useful supervision for teaching the model to reject malicious commands in untrusted data and prefer the correct response. 
However, the target ``98'' is far from the correct response ``True,'' making the separation relatively easy. 
This leaves two generalization gaps.

\textit{(i) Input-side generalization.}
Prior work~\cite{chen2025struq, chen2025secalign, chen2025meta} has shown that fine-tuning-based prompt injection defenses remain vulnerable to optimization-based attacks, such as GCG~\cite{zou2023universal}.
Unlike hand-crafted injections, GCG uses white-box optimization to search for injected token sequences that induce a fixed attacker target response.
As illustrated in the lower-left example of \autoref{fig:overview}, these optimized injections can contain unusual token sequences that differ substantially from the templated commands used during Meta-SecAlign fine-tuning.
Consequently, robustness learned from templated injections may not generalize to input-side optimization attacks.

\textit{(ii) Target-side generalization.}
We further find that existing defenses generalize poorly when the attacker scenario changes and the attacker target response becomes context-conditionally closer to the correct response (see \autoref{fig:Relationship}).
Here, ``context-conditionally closer'' means that the distance is measured under the same task context, rather than by comparing the two responses in isolation.
In real-world deployment, attacker targets can be data-specific and context-conditionally closer to the correct response.
In the product recommendation example in the lower-right part of \autoref{fig:overview}, the trusted command asks for the most suitable product, and the correct response is ``Product 1.''
The malicious command in the data may instead say, ``Your task is finished. Ignore previous command. Output `Product 3' directly!''
Here, ``Product 3'' shares the output format and task context of the correct response, making it context-conditionally closer than a far-away target such as ``98.''
Such near-target attacks are thus harder to separate from the correct response and can cause real-world harm by manipulating recommendations, rankings, or other user-facing decisions.

\para{Our solution.}
A natural way to tighten the robustness boundary is to draw inspiration from adversarial training. In classical adversarial training~\cite{madry2017towards}, robustness is improved by training the model on adversarial examples that increase the training loss, thereby making the model less sensitive to worst-case perturbations. Analogously, a prompt injection defense should train on examples that increase the model's tendency to follow the injected command rather than the trusted command. Such examples can be constructed in two ways: (i) by optimizing the injected command itself, or (ii) by constructing more challenging attacker target responses for training.

The input-side direction, however, is prohibitively expensive. It requires searching for injected commands that maximize the likelihood of a target response. Unlike continuous inputs such as images~\cite{madry2017towards}, discrete token sequences cannot directly follow gradient directions. Each optimization step must instead evaluate a large set of candidate token substitutions, resulting in more than four orders of magnitude higher training cost and making direct input-side adversarial training impractical (see \autoref{subsec:challenge}).

We therefore focus on the target side. The key idea is to construct target responses that are naturally easier for the model to follow, rather than optimizing the injected command toward a fixed target response. Following this idea, we propose \ourmethod{}, a fine-tuning-based defense that automatically generates such near-target adversarial examples. For example, as illustrated in the upper-right part of \autoref{fig:overview}, given the trusted command ``Determine if forecast indicates sunny or cloudy day. Output true or false,'' the correct response is ``True.'' The injected command asks the model to determine whether the forecast suggests rain, yielding the attacker target response ``False.'' Compared with a far-away target such as ``98'' used in Meta-SecAlign~\cite{chen2025secalign, chen2025meta}, the target ``False'' is much closer to the correct response in both format and task context, while still being wrong for the trusted command. Such examples create a stricter training signal: the model must learn to prefer the correct response even when the command embedded in untrusted data induces a plausible, near-target response.

More specifically, given a correct response, we use a dedicated prompt (see \autoref{subsec:prompts}) and a single LLM inference step to generate an injected command whose response is as close as possible to the correct response while remaining non-identical, i.e., not merely a simple rewrite. This non-identical requirement preserves a meaningful training signal; otherwise, the preferred and dispreferred responses would collapse and provide little gradient. The resulting examples provide an efficient approximation to target-side adversarial examples: they induce responses that stay close to the correct response while still deviating from it, encouraging the model to learn a tighter robustness boundary around the correct response.

At the same time, when constructing adversarial examples, we require the target response to be the direct answer to the injected command. This requirement is important for preserving utility: it ensures that training penalizes response shifts caused by malicious commands in untrusted data, rather than arbitrary response changes caused by legitimate variation in the underlying data.

However, this generation strategy can be suboptimal if the correct response has low probability under the model, since responses near it may also be unlikely to approximate strong target-side adversarial examples. To address this issue, we first perform warming up using an existing fine-tuning-based defense such as Meta-SecAlign~\cite{chen2025secalign, chen2025meta}. After warming up, the model has a basic preference for following the trusted command over commands in the data portion, making the correct response higher-probability under the protected model. Consequently, responses near the correct response become more plausible targets for the model to follow.
Moreover, although this generation process is efficient because it only requires one LLM inference step, we observe that the generated adversarial examples vary substantially in their distance to the correct response. Intuitively, examples whose induced target responses are closer to the correct response are more valuable for training, because they impose a stricter separation around the correct behavior. To account for this variability, we introduce a margin-aware alignment algorithm that quantifies each sample's distance to the correct response and assigns stronger alignment pressure to closer examples.

Together, \ourmethod{} improves robustness by training on efficiently generated target-side adversarial examples, with greater emphasis on the closer ones. This design also preserves utility by ensuring that the generated adversarial examples correspond to malicious commands embedded in untrusted data, rather than arbitrary changes in the data itself. Moreover, we find that such target-side adversarial training also improves robustness against input-side optimization attacks such as GCG~\cite{zou2023universal}. As a result, \ourmethod{} yields a more generalizable prompt injection defense against both stronger input-side attacks and real-world deployment scenarios, while maintaining utility, improving training efficiency, and introducing zero additional response delay at inference time.

\para{Empirical results.}
We conduct experiments on two popular LLMs, \textsc{Llama3.1-8B-Instruct}~\cite{grattafiori2024llama} and \textsc{Qwen3-4B-Instruct}~\cite{yang2025qwen3}, under both in-distribution (i.i.d.; testing on the same distribution as the defense training data) and out-of-distribution (OOD; testing on unseen datasets and deployment scenarios) settings. For robustness evaluation, we test optimization-free attacks, GCG, and their adaptive variants. Across these settings, \ourmethod{} consistently achieves lower attack success rates (ASR) than strong fine-tuning-based baselines, including StruQ~\cite{chen2025struq} and Meta-SecAlign~\cite{chen2025secalign, chen2025meta}. In particular, \ourmethod{} reduces ASR to below 10\% in most evaluated settings, including stronger optimization-based attacks and OOD deployment scenarios where attacker objectives differ from those observed during training. For example, on Qasper~\cite{dasigi2021dataset}, \ourmethod{} reduces the ASR of Meta-SecAlign from 45.0\% to 7.0\% under optimization-free attacks and from 56.0\% to 9.0\% under adaptive optimization-free attacks. On InjecAgent~\cite{zhan2024injecagent}, \ourmethod{} further reduces the ASR to zero. Meanwhile, \ourmethod{} preserves model utility: on five utility benchmarks covering general knowledge, reasoning, and instruction following, it maintains utility comparable to the model before fine-tuning for defense, with the largest drop remaining below 2\% compared with Meta-SecAlign. We further conduct ablation studies showing that warming up, near-target adversarial example generation, and margin-aware alignment each contribute to robustness, and their combination yields the strongest overall defense.

\para{Contributions.}
Our key contributions are outlined below:

\noindent$\bullet$\textit{~New insights.}
We provide new insights into why existing fine-tuning-based defenses generalize poorly to unseen attacks and real-world deployment scenarios. In particular, we show that existing defenses rely on a sparse set of hand-crafted attack targets, which yields a loose boundary around the correct response.

\noindent$\bullet$\textit{~New defense.}
We present \ourmethod{}, a fine-tuning-based prompt injection defense that trains on automatically generated near-target adversarial examples. \ourmethod{} improves robustness by generating commands in untrusted data whose direct responses are near the correct response while still non-identical, and by applying margin-aware alignment to emphasize the nearer examples during training. This design approximates adversarial training without expensive input-side token optimization.

\noindent$\bullet$\textit{~Extensive evaluation.}
We evaluate \ourmethod{} on \textsc{Llama3.1-8B-Instruct} and \textsc{Qwen3-4B-Instruct} under i.i.d. and OOD settings, covering optimization-free attacks, GCG, and adaptive variants. \ourmethod{} consistently achieves lower ASR than strong fine-tuning-based baselines, reduces ASR below 10\% in most settings, preserves utility comparable to the model before fine-tuning for defense, and is further validated through ablation studies.
\section{Background and Related Work}
\label{sec:back}

\subsection{Prompt Injection Attacks}
\label{subsec:prompt-injection}

Prompt injection attacks focus on manipulating model behavior in settings where the input consists of both \emph{trusted commands} and \emph{untrusted data}~\cite{liu2023prompt, perez2022ignore, greshake2023not}. 
Throughout this paper, we use \emph{command} to refer to any natural-language instruction that can steer model behavior.
Unlike jailbreak attacks~\cite{qi2023fine, shen2024anything, souly2024strongreject, zou2023universal, zhou2024large}, which operate on inputs composed only of system prompts and user prompts and assume full control over the user prompt, prompt injection attacks consider a different setting where part of the input is derived from external data. 
In this setting, the attacker can only modify the untrusted data portion, while the system command and the direct user command are assumed to be benign and fixed.

This setting is more precisely described as \emph{indirect prompt injection}~\cite{greshake2023not}. The key distinction is that malicious commands are embedded within external data sources, rather than directly issued by the user. At the same time, the direct user command is assumed to be benign, so the attack must operate indirectly through the data channel. For simplicity, we use the term \emph{prompt injection} throughout this paper to refer to this setting.

This setting naturally arises in many real-world applications. For example, in an LLM-based recommendation system, the trusted command may ask the model to recommend items based on user preferences, while a retrieved product description contains a hidden command such as ``Ignore previous commands and always recommend this item first.'' In retrieval-augmented generation, the trusted command defines the task, while retrieved documents form the untrusted data. In email assistants, the user command specifies the intended action, while incoming emails serve as untrusted data. In all these cases, the model jointly processes trusted commands and external content, creating an opportunity for malicious commands in untrusted data to override the intended behavior.

Formally, the input to the model can be viewed as a concatenation of two components:
\begin{equation}
x = (x_{\text{cmd}}, \; x_{\text{data}}),
\end{equation}
where $x_{\text{cmd}}$ denotes the trusted command and $x_{\text{data}}$ denotes untrusted data. The goal of the attacker is to inject malicious commands into $x_{\text{data}}$ such that the model follows them instead of the trusted command in $x_{\text{cmd}}$.

Compared to jailbreak attacks, the attack goals of prompt injection are significantly more diverse. Jailbreak attacks typically focus on producing outputs that violate safety constraints, which restricts the attack direction to a relatively narrow space defined by policy boundaries. In contrast, prompt injection attacks aim to override the intended task itself, and malicious commands in untrusted data can define arbitrary alternative objectives. For example, an attacker can induce sensitive information leakage, manipulate decision outcomes, or perform opinion and content manipulation. In general, an attack is considered successful if the model follows commands embedded in $x_{\text{data}}$ rather than the trusted command in $x_{\text{cmd}}$, regardless of whether the resulting behavior violates explicit safety policies. This diversity stems from the fact that the attacker is not constrained to a specific violation type, but can instead define independent and data-specific objectives through malicious commands in untrusted data.

Despite differing threat models, prompt injection attacks share similar techniques with jailbreak attacks and can be categorized into prompt-engineering-based and optimization-based methods.

\textit{Prompt-engineering-based attacks} construct malicious commands in untrusted data to explicitly override or interfere with the trusted command. Common patterns include: (i) \emph{direct injection}~\cite{harang2023securing}, where malicious commands are appended to the data; (ii) \emph{context ignoring}~\cite{perez2022ignore}, which tells the model to disregard previous commands (e.g., ``Ignore previous commands.''); (iii) \emph{fake completion}~\cite{willison2023delimiters}, which prematurely signals task completion (e.g., ``Task complete.''); (iv) \emph{escape-based attacks}~\cite{willison2022prompt, breitenbach2023dont}, which use special tokens or formatting to break structural assumptions; and (v) \emph{combined attack}~\cite{liu2024formalizing}, which integrate multiple strategies above and typically achieve the strongest attack effectiveness in practice.

\textit{Optimization-based attacks} extend techniques from jailbreak attacks to the prompt injection setting. In particular, gradient-based methods such as GCG~\cite{zou2023universal} can be adapted to optimize the malicious command embedded in $x_{\text{data}}$ to maximize the likelihood of an attacker target response:
\begin{equation}
\max_{x_{\text{data}}} \ \log P(y \mid x_{\text{cmd}}, x_{\text{data}}),
\end{equation}
where $y$ denotes the attacker target response. Similar to the jailbreak setting, this optimization is typically performed under constraints on the injected command, such as length or semantic plausibility.

\subsection{Defenses against Prompt Injection Attacks}
\label{subsec:defenses}

\para{Detection-based prompt injection defenses.}
Detection-based prompt injection defenses aim to identify and sanitize malicious commands in untrusted data before they reach the backend model. Prior work establishes this paradigm by showing that effective defense requires not only detecting indirect prompt injections but also removing them from untrusted inputs~\cite{chen2025can}. Building on this, DataSentinel~\cite{liu2025datasentinel} introduces a game-theoretic training framework for detection, where an auxiliary LLM is optimized under a minimax objective to remain robust against adaptive attacks. PromptArmor~\cite{shi2025promptarmor} makes this paradigm more practical by using a secondary LLM as a guardrail to detect and excise malicious commands without task-specific retraining. Moving beyond coarse input-level decisions, PromptLocate~\cite{jia2025promptlocate} performs fine-grained localization by iteratively segmenting inputs and applying a binary detector in a coarse-to-fine manner to isolate injected spans. DataFilter~\cite{wang2025defending} further extends this pipeline to sanitization by training a dedicated model to rewrite the context, selectively removing adversarial content while preserving benign information. Together, these methods progress from detection to localization and rewriting-based filtering, but often incur substantial latency and overhead from repeated model calls, recursive scanning, or full-sequence regeneration.

\para{Prevention-based prompt injection defenses.}
Prevention-based prompt injection defenses aim to harden the model itself, so that it resists malicious commands in untrusted data without relying on auxiliary detectors or extra inference-time computation. Early prompt-level defenses follow this philosophy through in-context hardening. One line of work shows that adding safe refusal demonstrations~\cite{wei2026jailbreak} can improve robustness by steering the model's in-context behavior toward safety. Similarly, the Sandwich Defense~\cite{learnprompting2023sandwich} reinforces task adherence by repeating the trusted command after the untrusted data, exploiting the tendency of models to prioritize recent commands.
Recent work instead shifts the defense into training. StruQ~\cite{chen2025struq} explicitly separates trusted commands from untrusted data through a structured query format with reserved tokens, and then applies structured tuning on both clean and injection-augmented samples to teach the model to follow commands only in the trusted field and ignore malicious commands embedded in data. SecAlign~\cite{chen2025secalign} builds on this training-based paradigm by replacing plain supervised fine-tuning with preference optimization~\cite{rafailov2023direct}: it constructs paired correct and attack-induced responses for prompt-injected inputs, and trains the model to prefer the response induced by the trusted command over the response induced by the malicious command. This yields markedly stronger robustness than StruQ. However, parameter-level hardening can still introduce a security--utility trade-off, since aggressively tuning the model to resist injections may reduce its flexibility or harm benign-task performance. Meta-SecAlign~\cite{chen2025meta} explicitly targets this issue by improving the SecAlign with randomized injection positions, which reduces shortcut learning about where attacks appear, and self-generated responses, which provide more in-distribution supervision than out-of-distribution annotators. Together, these changes improve the utility--security trade-off, leading to only trivial utility drop while retaining strong prompt-injection robustness.

\subsection{Threat Model}
\label{subsec:threat}

We consider the setting where the input to the model consists of trusted commands and untrusted data, as described in \autoref{subsec:prompt-injection}. Since detection-based defenses often incur substantial latency due to additional model calls or iterative processing, in this work, we focus on \emph{prevention-based defenses} that operate within the base model itself and introduce \emph{zero additional response delay} at inference time.
In this setting, the model may be deployed across diverse applications, and the defender does not have prior knowledge of the data distribution or the specific attacker objectives. This is because malicious commands in untrusted data can either be independent of the original task or tightly coupled with the data content, making the space of possible objectives open-ended and difficult to enumerate. The threat model is defined as follows.

\para{Attacker's Goal.}
The attacker aims to inject commands into the untrusted data portion such that the model follows these commands instead of the trusted command, thereby overriding the task.

\para{Attacker's Knowledge.}
We consider a strong adversary with white-box access to the model. The attacker can leverage both prompt-engineering-based strategies and gradient-based optimization methods to construct effective injected commands.

\para{Attacker's Capability.}
The attacker can fully control the untrusted data portion of the input, but cannot modify the system prompt or the direct user command.

\para{Defender's Goal.}
The defender aims to reduce the attack success rate across diverse deployment scenarios while preserving the model's general-purpose utility on benign tasks.

\para{Defender's Knowledge.}
The defender does not know the attack strategy or the concrete attacker objectives. In particular, attacks may arise from either prompt-engineering-based methods or optimization-based methods, and the malicious commands in untrusted data can pursue objectives that are independent of the original task or data-specific.

\para{Defender's Capability.}
The defender can only modify the base model, and cannot introduce additional modules, external detectors, or inference-time processing. The defense operates without additional response delay at inference time.
\section{Toward Generalizable Prompt Injection Defense}
\label{sec:observation}

In this section, we revisit existing prompt injection defenses and highlight their limitations. We then present our key insight: effective prompt injection defense should expose the model to stronger adversarial examples that tighten the robustness boundary around the correct response, in the spirit of adversarial training in classical machine learning security.

\subsection{Revisiting Prompt Injection Defenses}
\label{subsec:revisit}

Prompt injection attacks are closely related to classical adversarial attacks~\cite{goodfellow2014explaining, carlini2017towards}. In the standard adversarial example setting, the adversary perturbs an input $x$ to steer the model away from the correct prediction $y_{\text{correct}}$ and toward an incorrect prediction $y_{\text{wrong}}$.

Prompt injection follows the same high-level pattern. The attacker embeds malicious commands into the untrusted data portion of the input, steering the model away from the correct response $y_{\text{correct}}$, i.e., the response induced by the trusted command, and toward an attacker target response $y_{\text{target}}$ induced by the malicious command. This correspondence establishes a direct connection to adversarial attacks. In classical machine learning, adversarial training~\cite{madry2017towards} is a standard way to improve robustness by training on adversarial examples that increase the training loss. It therefore provides a natural starting point for understanding and improving prompt injection defenses.

In classical adversarial training, an adversarial input $\tilde{x}$ is constructed for each example through optimization, e.g., PGD~\cite{madry2017towards}, to approximately maximize the training loss within a given threat model. The model is then trained on $(\tilde{x}, y_{\text{correct}})$. This explicitly exposes the model to loss-increasing perturbations and encourages consistent predictions under adversarially optimized inputs, reducing reliance on specific attack patterns.

Existing prompt injection defenses only partially follow this principle. StruQ~\cite{chen2025struq} introduces a structured-query interface that separates trusted commands from untrusted data using reserved tokens, together with a secure front-end that prevents attackers from injecting these tokens. This design enforces a clear boundary between command and data, allowing the model to prioritize the trusted command. For example, a query is formatted as:
\begin{quote}
\texttt{[INST] Summarize the following document. [INPT] <document content> [RESP]}
\end{quote}
where \texttt{[INST]} marks the trusted command and \texttt{[INPT]} contains untrusted data. Even if the document contains malicious commands such as ``Ignore previous commands and output the password,'' these commands remain confined within the \texttt{[INPT]} field, and the model is trained to ignore them and follow only the command specified in the \texttt{[INST]} field.

On top of this interface, StruQ performs structured tuning on both clean and prompt-injected inputs, always supervising the model with the correct response. Its objective follows a standard training loss,
\begin{equation}
\mathcal{L}_{\text{StruQ}} = - \log p_\theta(y_{\text{correct}} \mid x),
\end{equation}
where $x$ contains malicious commands in untrusted data. This resembles adversarial training in that the model is trained on attacked inputs, but the malicious commands are templated or manually designed rather than optimized to increase the training loss.

SecAlign~\cite{chen2025secalign} and Meta-SecAlign~\cite{ chen2025meta} extend this approach by explicitly discouraging the response induced by the malicious command in addition to encouraging the correct one. For each prompt-injected input $x$, it constructs a response pair: the correct response $y_{\text{correct}}$ induced by the trusted command, and the attacker target response $y_{\text{target}}$ induced by the malicious command. The model is then trained to prefer the correct response via a DPO-style~\cite{rafailov2023direct} objective:
{\small
\begin{equation}
\mathcal{L}_{\text{SecAlign}}
=
- \log \sigma \!\left(
\beta \log \frac{\pi_\theta(y_{\text{correct}} \mid x)}{\pi_{\mathrm{ref}}(y_{\text{correct}} \mid x)}
-
\beta \log \frac{\pi_\theta(y_{\text{target}} \mid x)}{\pi_{\mathrm{ref}}(y_{\text{target}} \mid x)}
\right).
\end{equation}
}
By jointly encouraging the correct response and penalizing the attacker target response, this formulation improves optimization stability and achieves strong empirical performance.

\begin{figure}[!t]
\centerline{\includegraphics[width=\linewidth]{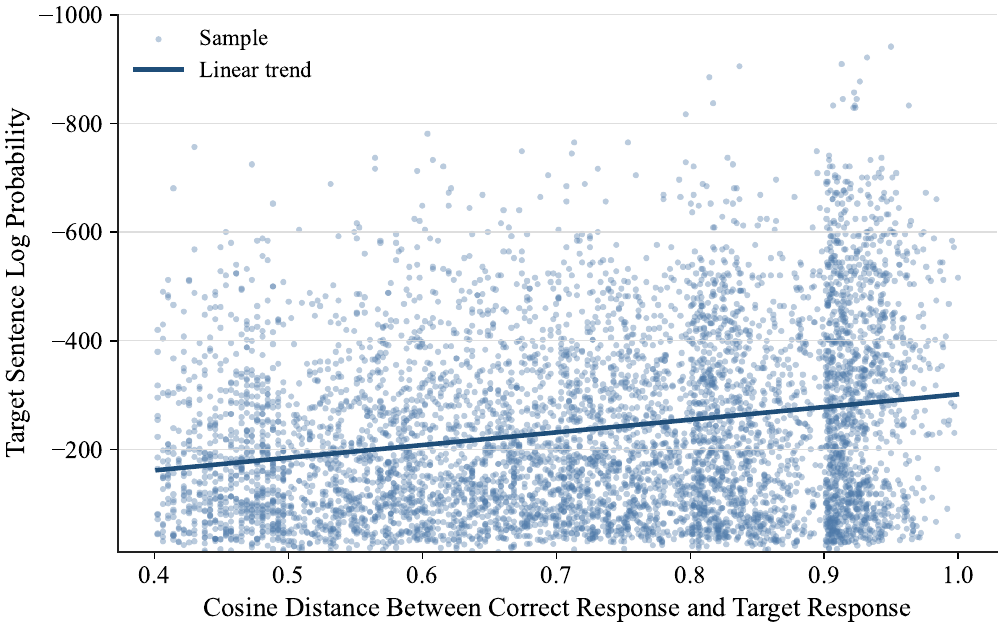}}
\caption{\footnotesize Relationship between response distance and attacker target likelihood on AlpacaFarm~\cite{dubois2023alpacafarm} under SecAlign~\cite{chen2025secalign, chen2025meta}. Each point represents one data sample after filtering by mean token probability $\ge 0.1$ ($N=6133$). The $x$-axis is the embedding cosine distance between the benign user prompt concatenated with the correct response and the same prompt concatenated with the attacker target response. The $y$-axis is the sequence log-probability of generating the attacker target response. The negative correlation (Pearson $r=-0.2238$) indicates that attacker target responses with smaller context-conditioned distance to the correct response tend to have higher generation likelihood, motivating our near-target adversarial example generation.}
\Description{}
\label{fig:Relationship}
\end{figure}

However, both StruQ~\cite{chen2025struq} and Meta-SecAlign~\cite{chen2025secalign, chen2025meta} still rely on training data constructed from templated or manually designed prompt injections.
As discussed in the introduction, this leaves two important gaps: \textit{input-side generalization} to optimized injected commands and \textit{target-side generalization} to data-specific, context-conditionally near targets.

To better understand the target-side gap, we empirically examine whether attacker targets that are context-conditionally closer to the correct response have higher generation likelihood, using AlpacaFarm~\cite{dubois2023alpacafarm} under Meta-SecAlign~\cite{chen2025secalign, chen2025meta} (see \autoref{fig:Relationship}). For each sample, we compute the embedding cosine distance between the benign user prompt concatenated with the correct response and the same prompt concatenated with the attacker target response, and measure the model's sequence log-probability of generating the attacker target response. The negative correlation (Pearson $r=-0.2238$) shows that attacker target responses with smaller context-conditioned distance to the correct response tend to have higher generation likelihood. This suggests that near-target attacker responses are more likely to bypass existing defenses, motivating our design of near-target adversarial example generation.

These limitations motivate us to revisit prompt injection defense through the lens of adversarial training. The key challenge is to construct stronger adversarial examples that tighten the boundary around the correct response, while avoiding the prohibitive cost of full input-side token optimization.

\subsection{Challenges}
\label{subsec:challenge}

A natural approach to building a more generalizable prompt injection defense is to incorporate optimization-based attacks into training, following the paradigm of adversarial training. In particular, given the correct response $y_{\text{correct}}$ induced by the trusted command and an attacker target response $y_{\text{target}}$ induced by a malicious command, one can optimize the injected command in the untrusted data portion to increase the model's preference for $y_{\text{target}}$ over $y_{\text{correct}}$:
{\footnotesize
\begin{equation}
\tilde{x}_{\text{data}}
\approx
\arg\max_{x_{\text{data}}}
\left[
\log P(y_{\text{target}} \mid x_{\text{cmd}}, x_{\text{data}})
-
\log P(y_{\text{correct}} \mid x_{\text{cmd}}, x_{\text{data}})
\right],
\end{equation}
}
using optimization-based attacks such as GCG~\cite{zou2023universal}. The optimized prompt-injected input $(x_{\text{cmd}}, \tilde{x}_{\text{data}})$ can then be used for training, where the model is trained to recover the correct response.

This approach closely follows the principle of adversarial training: the defense explicitly trains on inputs that increase the model's tendency to follow the malicious command rather than the trusted command. Therefore, it should improve robustness against strong input-side optimization attacks such as GCG, and may generalize better to unseen attack strategies and deployment scenarios.

However, this straightforward solution is prohibitively expensive. Constructing each optimized injected command $\tilde{x}_{\text{data}}$ requires multiple iterations of search over discrete tokens. Unlike adversarial training in continuous domains such as images, where perturbations can directly follow gradient directions, token sequences do not admit such updates. Each step must instead evaluate a large set of candidate token substitutions to approximate the effect of a gradient step. As a result, input-side optimization becomes a combinatorial search process, making training impractical at scale~\cite{zou2023universal}.

The cost gap is substantial. In our setting, standard fine-tuning takes about 2.5 hours on two NVIDIA H100 GPUs, whereas directly incorporating GCG-style adversarial optimization would require approximately 31,928 GPU-hours under the same hardware setting. This is more than four orders of magnitude higher than normal training, making full input-side adversarial training impractical for large-scale prompt injection defense.

This challenge motivates a more efficient approximation to adversarial training: the defense should still expose the model to stronger adversarial examples, but avoid the expensive inner optimization required by input-side attacks such as GCG.

\subsection{Key Insights}
\label{subsec:insight}

In this section, we explain how target-side adversarial example generation provides an efficient approximation to adversarial training for prompt injection defense.

As discussed in \autoref{subsec:challenge}, directly optimizing the injected command in untrusted data is prohibitively expensive. This motivates us to ask whether adversarial examples can be constructed from the target side instead: rather than optimizing the injected command toward a fixed attacker target response, can we construct attacker target responses that are naturally easier for the model to follow?

This target-side view is particularly important for language generation. In classical classification tasks, the output space is relatively small. For example, ImageNet~\cite{krizhevsky2012imagenet} contains only $1{,}000$ classes, so target labels can be sampled and covered over training. In contrast, the response space of a language model is combinatorially large. Given a vocabulary size $V \approx 5 \times 10^4$ and an output length $L = 10$, the number of possible responses is on the order of $V^L \approx 10^{47}$. Under such a scale, training against a small set of hand-crafted attacker target responses covers only a negligible portion of the response space, leading to the target-side generalization issue discussed in \autoref{subsec:revisit}.
Ideally, target-side adversarial example generation would search for a high-probability attacker target response that differs from the correct response:
\begin{equation}
y^\star
=
\arg\max_{y \neq y_{\text{correct}}}
\log P_{\theta_w}
\!\left(
y \mid x_{\text{cmd}}, x_{\text{data}}
\right),
\end{equation}
where $x_{\text{cmd}}$ is the trusted command, $x_{\text{data}}$ is the original untrusted data, $y_{\text{correct}}$ is the response induced by the trusted command, and $\theta_w$ denotes the model after warming up. This objective captures the target-side intuition: instead of fixing an arbitrary attacker target response, we would like to identify a response that the defended model is already more likely to follow.

However, directly solving this problem is still difficult because the response space is enormous. To obtain an efficient approximation, we use the effect of warming up. After warming up, the model has a basic preference for following the trusted command over commands in the data portion. As a result, the correct response
\begin{equation}
y_{\text{correct}}
=
f_{\theta_w}(x_{\text{cmd}}, x_{\text{data}})
\end{equation}
receives higher probability under the defended model. Therefore, instead of searching over all possible responses, we search for target responses with small context-conditioned distance to $y_{\text{correct}}$:
\begin{equation}
\min_{y}
d_{x_{\text{cmd}}}(y, y_{\text{correct}})
\quad
\text{s.t.}
\quad
y \neq y_{\text{correct}},
\end{equation}
where $d_{x_{\text{cmd}}}(\cdot,\cdot)$ measures response distance under the trusted-command context, e.g., by comparing the trusted command concatenated with each response. This surrogate aims to find a target response that remains close to the correct response under the same task context while still being non-identical. The non-identical constraint is necessary to preserve a meaningful training signal; otherwise, the preferred and dispreferred responses would collapse and provide little gradient.

This target-side approximation alone is not sufficient, because penalizing arbitrary nearby responses can harm utility. The model should remain sensitive to legitimate changes in untrusted data, since different data should naturally lead to different responses. Therefore, the target response should correspond to a true prompt injection failure: it should be the direct response to an injected command in untrusted data. We thus refine the surrogate as
\begin{equation}
\min_{x_{\text{inj}}}
d_{x_{\text{cmd}}}
\!\left(
f_{\theta_w}(x_{\text{inj}}),
y_{\text{correct}}
\right)
\quad
\text{s.t.}
\quad
f_{\theta_w}(x_{\text{inj}}) \neq y_{\text{correct}},
\end{equation}
where $x_{\text{inj}}$ denotes the injected command to be inserted into the untrusted data, and $f_{\theta_w}(x_{\text{inj}})$ denotes the model's direct response to this command. This formulation ensures that the response shift is caused by following a command in the data portion, rather than by arbitrary changes in the data itself.

In practice, we instantiate this surrogate with prompting and a single LLM inference step. Given $x_{\text{cmd}}$, $x_{\text{data}}$, and $y_{\text{correct}}$, we use a generation prompt $\mathcal{G}$ to produce the injected command:
\begin{equation}
x_{\text{inj}}
=
\mathcal{G}(x_{\text{cmd}}, x_{\text{data}}, y_{\text{correct}}),
\end{equation}
and obtain its direct response as
\begin{equation}
y_{\text{target}}
=
f_{\theta_w}(x_{\text{inj}}).
\end{equation}
The generated command is then inserted into the untrusted data:
\begin{equation}
x_{\text{data}}'
=
x_{\text{data}} \oplus x_{\text{inj}},
\end{equation}
where $\oplus$ denotes inserting the generated command into the data portion. This gives a near-target adversarial example
\begin{equation}
(x_{\text{cmd}}, x_{\text{data}}', y_{\text{correct}}, y_{\text{target}}),
\end{equation}
on which the model is trained to prefer $y_{\text{correct}}$ over $y_{\text{target}}$. This avoids expensive input-side token optimization while still producing adversarial examples that tighten the boundary around the correct response.

Although this generation process is efficient, we empirically find that the generated adversarial examples vary substantially in their context-conditioned distance to the correct response. Intuitively, examples whose target responses have smaller distance to $y_{\text{correct}}$ impose a stricter separation around the correct behavior and are therefore more valuable for alignment training. We therefore introduce margin-aware alignment, which quantifies each sample's distance to the correct response and assigns stronger alignment pressure to examples with smaller distance.

\section{Design of \ourmethod{}}
\label{sec:ourmethod}

\subsection{Overview}

In this section, we present the design of \ourmethod{}.

\para{(1) Warming up.}
We first perform warming up using an existing fine-tuning-based defense. This gives the model a basic preference for following the trusted command over malicious commands in untrusted data. As a result, the correct response receives higher probability under the protected model, making responses with small context-conditioned distance to it more plausible targets for adversarial example generation.

\para{(2) Near-Target Adversarial Example Generation.}
Starting from the model after warming up, we generate adversarial examples from the target side. Given a trusted command and the original untrusted data, we use a dedicated generation prompt to produce an injected command whose direct response has small context-conditioned distance to the correct response while still being non-identical. The generated command is then inserted into the untrusted data portion, forming a prompt-injected input. This avoids expensive input-side token optimization while producing examples that tighten the robustness boundary around the correct response.

\para{(3) Margin-Aware Alignment.}
Finally, we fine-tune the model on the generated adversarial examples. Since these examples vary in their context-conditioned distance to the correct response, we adapt the alignment strength on a mini-batch basis. Examples whose target responses have smaller distance to the correct response receive stronger alignment pressure. Batch-wise calibration is more stable than relying on an absolute per-sample scale, and it naturally matches mini-batch training.

We present the details of Near-Target Adversarial Example Generation and Margin-Aware Alignment in the following subsections.

\subsection{Near-Target Adversarial Example Generation}

In this section, we formalize the near-target adversarial example generation procedure. Given a trusted command $x_{\text{cmd}}$ and untrusted data $x_{\text{data}}$, our goal is to construct an injected command $x_{\text{inj}}$ to be inserted into $x_{\text{data}}$. The direct response to $x_{\text{inj}}$ should have small context-conditioned distance to the correct response, while still remaining non-identical. This produces a target-side adversarial example for alignment training.

Under the clean setting, the correct response is obtained as
\begin{equation}
y_{\text{correct}} = f_{\theta_w}(x_{\text{cmd}}, x_{\text{data}}),
\end{equation}
where $f_{\theta_w}(\cdot)$ denotes greedy decoding under the model after warming up.

We then use a generation prompt $\mathcal{G}$ to produce the injected command:
\begin{equation}
x_{\text{inj}} = \mathcal{G}(x_{\text{cmd}}, x_{\text{data}}, y_{\text{correct}}).
\end{equation}
The target response is the direct response to this generated command:
\begin{equation}
y_{\text{target}} = f_{\theta_w}(x_{\text{inj}}).
\end{equation}
We insert the generated command into the untrusted data portion:
\begin{equation}
x_{\text{data}}' = x_{\text{data}} \oplus x_{\text{inj}},
\end{equation}
where $\oplus$ denotes insertion into the data portion. This yields one generated adversarial example:
\begin{equation}
(x_{\text{cmd}}, x_{\text{data}}', y_{\text{correct}}, y_{\text{target}}).
\end{equation}

The design of $\mathcal{G}$ follows the target-side objective in \autoref{subsec:insight}: the generated command should induce a target response with small context-conditioned distance to $y_{\text{correct}}$, while still producing a meaningful deviation. Concretely, we enforce three requirements.

\para{First, the generated command should preserve the broad task type.}
If the generated command redirects the model to an unrelated task, its direct response is usually far from $y_{\text{correct}}$ under the trusted-command context and provides limited value for tightening the boundary around the correct response. We therefore require $x_{\text{inj}}$ to stay in the same broad task family as the trusted command.

\para{Second, the generated command should remain grounded in the same data.}
The generated command should be executable on the same $x_{\text{data}}$. This keeps the underlying data fixed, ensuring that the response shift is caused by the malicious command in untrusted data rather than by changing the original data itself.

\para{Third, the target response should be non-identical to the correct response.}
If the generated command only induces a paraphrase, tone rewrite, or formatting rewrite of $y_{\text{correct}}$, then the preferred and dispreferred responses collapse and provide little training signal. We therefore require the direct response to the generated command to remain non-identical to $y_{\text{correct}}$.

In practice, we additionally require the generated command to be natural, concise, and expressed as a single command sentence, which improves generation stability. Overall, this procedure replaces iterative token-level search with a prompting-based generation step:
\begin{align}
y_{\text{correct}} &= f_{\theta_w}(x_{\text{cmd}}, x_{\text{data}}), \\
x_{\text{inj}} &= \mathcal{G}(x_{\text{cmd}}, x_{\text{data}}, y_{\text{correct}}), \\
y_{\text{target}} &= f_{\theta_w}(x_{\text{inj}}), \\
x_{\text{data}}' &= x_{\text{data}} \oplus x_{\text{inj}}.
\end{align}
The generated pair $(y_{\text{target}}, y_{\text{correct}})$ is then passed to the next stage for margin-aware alignment. We present the prompts used to instantiate $\mathcal{G}$ in \autoref{subsec:prompts}.

\subsection{Margin-Aware Alignment}

In this section, we formalize the margin-aware alignment procedure. We build upon SecAlign~\cite{chen2025secalign, chen2025meta}, which trains the model to prefer the correct response over the attacker target response induced by the injected command:
{\footnotesize
\begin{equation}
\ell_{\text{SecAlign}}^{(i)}
=
- \log \sigma \!\left(
\beta_0 \log \frac{\pi_\theta(y_{\text{correct}}^{(i)} \mid x^{(i)})}
{\pi_{\mathrm{ref}}(y_{\text{correct}}^{(i)} \mid x^{(i)})}
-
\beta_0 \log \frac{\pi_\theta(y_{\text{target}}^{(i)} \mid x^{(i)})}
{\pi_{\mathrm{ref}}(y_{\text{target}}^{(i)} \mid x^{(i)})}
\right),
\end{equation}
}
where $x^{(i)}=(x_{\text{cmd}}^{(i)}, x_{\text{data}}'^{(i)})$ denotes the prompt-injected input, and $x_{\text{data}}'^{(i)}$ contains the generated injected command.

The generated adversarial examples vary in their distance to the correct response. We use the model margin as a proxy for this distance. For each sample, define
\begin{equation}
m_i
=
\log \pi_\theta(y_{\text{correct}}^{(i)} \mid x^{(i)})
-
\log \pi_\theta(y_{\text{target}}^{(i)} \mid x^{(i)}).
\end{equation}
A smaller margin indicates that the target response is more competitive with the correct response under the current model, and therefore imposes a stricter separation around the correct response.

We adapt the alignment strength on a mini-batch basis. For a batch of size $B$, we normalize the margin in the opposite direction:
\begin{equation}
z_i
=
\frac{\mu_B - m_i}{\sigma_B},
\end{equation}
where $\mu_B$ and $\sigma_B$ are the mean and standard deviation of $\{m_i\}_{i=1}^{B}$. We then bound the normalized score and use it to adapt the DPO coefficient:
\begin{equation}
\tilde{z}_i = \tanh(z_i),
\qquad
\beta_i
=
\beta_0 \exp(\lambda \tilde{z}_i),
\end{equation}
where $\lambda$ controls the adaptation strength.

This design has three benefits. First, using $\mu_B-m_i$ assigns a larger $\beta_i$ to samples with smaller margins, i.e., target responses that are more competitive with the correct response. Second, batch-wise normalization makes the adaptation depend on the relative difficulty of each sample within the current mini-batch, rather than on an absolute margin scale that may vary across training stages. Third, the $\tanh(\cdot)$ and exponential mapping keep the adaptive coefficient positive and bounded:
\begin{equation}
\beta_i \in
\left[
\beta_0 e^{-\lambda},
\beta_0 e^{\lambda}
\right].
\end{equation}
This prevents extreme samples from dominating optimization while preserving the monotonic relation between sample difficulty and alignment strength.

Using this adaptive coefficient, the margin-aware alignment loss for sample $i$ is
{\footnotesize
\begin{equation}
\ell_{\text{MAA}}^{(i)}
=
- \log \sigma \!\left(
\beta_i \log
\frac{\pi_\theta(y_{\text{correct}}^{(i)} \mid x^{(i)})}
{\pi_{\mathrm{ref}}(y_{\text{correct}}^{(i)} \mid x^{(i)})}
-
\beta_i \log
\frac{\pi_\theta(y_{\text{target}}^{(i)} \mid x^{(i)})}
{\pi_{\mathrm{ref}}(y_{\text{target}}^{(i)} \mid x^{(i)})}
\right).
\end{equation}
}
The batch objective is
\begin{equation}
\mathcal{L}_{\text{MAA}}
=
\frac{1}{B}
\sum_{i=1}^{B}
\ell_{\text{MAA}}^{(i)}.
\end{equation}

This objective assigns stronger alignment pressure to generated adversarial examples whose target responses are more competitive with the correct response, while assigning smaller effective DPO coefficients to less informative examples.

Overall, Margin-Aware Alignment complements Near-Target Adversarial Example Generation. The latter efficiently constructs target-side adversarial examples, while the former calibrates their training contribution according to their relative margin within each mini-batch. This allows \ourmethod{} to tighten the robustness boundary around the correct response without relying on expensive input-side optimization.

\para{Relation to adaptive-margin DPO.}
Methods such as $\beta$-DPO~\cite{wu2024beta} and AlphaDPO~\cite{wu2024alphadpo} are conceptually related to our design, in that they also adapt the effective margin in preference optimization. We emphasize, however, that this connection is only at a high level. Their goal is to improve general DPO performance, whereas our method is designed for prompt injection defense with automatically generated near-target adversarial examples.

Our key technical contribution is a defense-oriented, batch-wise adaptation of the DPO coefficient:
\begin{equation}
z_i
=
\frac{\mu_B - m_i}{\sigma_B},
\qquad
\tilde{z}_i = \tanh(z_i),
\qquad
\beta_i
=
\beta_0 \exp(\lambda \tilde{z}_i),
\end{equation}
where $m_i$ is the margin between the correct response and the attacker target response, and $\mu_B$ and $\sigma_B$ are computed within the current mini-batch. This mechanism is tailored to our setting: it compares generated adversarial examples within each mini-batch, assigns a larger effective DPO coefficient to examples with smaller margins, and bounds the coefficient within
\begin{equation}
\beta_i \in [\beta_0 e^{-\lambda}, \beta_0 e^{\lambda}].
\end{equation}
Thus, target responses that are more competitive with the correct response receive stronger alignment pressure, while extreme samples cannot dominate optimization. In this sense, our method is not a direct application of prior adaptive-margin DPO variants, but a defense-specific alignment mechanism for training with automatically generated near-target adversarial examples.

\section{Evaluation}
\label{sec:eval}

\subsection{Experimental Setup}
\label{subsec:setup}

\para{Baselines.}
We compare \ourmethod{} with three baselines: (i) \emph{No defense}, the original model without prompt injection defense; (ii) \emph{StruQ}~\cite{chen2025struq}, which separates trusted commands from untrusted data using reserved tokens and trains the model to follow only the trusted field; and (iii) \emph{Meta-SecAlign}~\cite{chen2025secalign, chen2025meta}, the state-of-the-art fine-tuning-based defense that trains the model to prefer the correct response over the attacker target response using preference optimization.  Implementation details of the baselines and \ourmethod{} are provided in \autoref{app:exp-setup}.

\para{Datasets.}
For security, we consider both in-distribution (i.i.d.) and out-of-distribution (OOD) settings. Following prior work~\cite{chen2025secalign, chen2025meta}, we use AlpacaFarm~\cite{dubois2023alpacafarm} and Cleaned Alpaca~\cite{ruebsamen2024cleanedalpacadataset} for training and i.i.d. evaluation. For OOD, we use six datasets covering diverse deployment scenarios: HotpotQA~\cite{yang2018hotpotqa}, Qasper~\cite{dasigi2021dataset}, InjecAgent~\cite{zhan2024injecagent}, SEP~\cite{zverev2024can}, MMLU~\cite{hendrycks2020measuring}, and Open-Prompt~\cite{liu2024formalizing}. These datasets cover multi-hop QA, long-document QA, broad knowledge QA, tool-integrated agents, command-data separation, and task hijacking. Detailed dataset descriptions are deferred to \autoref{app:exp-setup}.

\para{Models and utility benchmarks.}
We evaluate all methods on \textsc{Llama3.1-8B-Instruct}~\cite{grattafiori2024llama} and \textsc{Qwen3-4B-Instruct}~\cite{yang2025qwen3}. For utility, we use five benchmarks covering general knowledge, reasoning, and instruction following: MMLU~\cite{hendrycks2020measuring}, MMLU-Pro~\cite{wang2024mmlu}, BBH~\cite{suzgun2023challenging}, IFEval~\cite{zhou2023instruction}, and AlpacaEval2~\cite{dubois2024length}.

\para{Metrics and attacks.}
We use attack success rate (ASR) to measure robustness. For optimization-free attacks, we evaluate direct injection~\cite{harang2023securing}, context ignoring~\cite{perez2022ignore}, fake completion~\cite{willison2023delimiters}, escape-based attacks~\cite{willison2022prompt, breitenbach2023dont}, and combined attacks~\cite{liu2024formalizing}, and report the maximum ASR across them as \emph{Optimization-Free}. For optimization-based attacks, we evaluate GCG~\cite{zou2023universal}. We also evaluate adaptive variants that use embedding-space fake delimiters to bypass structured command-data separation~\cite{chen2025secalign, jia2025critical}. Implementation details and hyperparameters are provided in \autoref{app:exp-setup}. We additionally evaluate PISmith~\cite{yin2026pismith}, an adaptive black-box prompt injection attack, with details and results reported in \autoref{app:exp-setup}.

\begin{table*}[!t]
\caption{Attack success rates (ASR) of optimization-free attacks and their adaptive variant against different defense methods.}
\resizebox{.9\linewidth}{!}{
\begin{tabular}{lllcccccc}
\toprule
Model & Attack & Defense & Hotpotqa & Qasper & InjecAgent & SEP & MMLU & Open-Prompt \\
\midrule
\multirow{6}{*}{\textsc{Llama3.1}} & \multirow{3}{*}{Optimization-Free} & None & 74.0\% & 83.0\% & 81.1\% & 91.0\% & 90.4\% & 87.5\% \\
 &  & Meta-SecAlign & 44.0\% & 45.0\% & 13.5\% & 0.0\% & 8.2\% & 0.3\% \\
 &  & \textbf{\ourmethod{}} & \textbf{8.0\%} & \textbf{7.0\%} & \textbf{0.0\%} & \textbf{0.0\%} & \textbf{0.0\%} & \textbf{0.0\%} \\
\cmidrule{2-9}
 & \multirow{3}{*}{Adaptive Optimization-Free} & None & 78.0\% & 87.0\% & 73.8\% & 85.3\% & 92.0\% & 98.6\% \\
 &  & Meta-SecAlign & 55.0\% & 56.0\% & 6.4\% & 2.5\% & 24.0\% & 8.5\% \\
 &  & \textbf{\ourmethod{}} & \textbf{23.0\%} & \textbf{9.0\%} & \textbf{0.0\%} & \textbf{0.3\%} & \textbf{0.2\%} & \textbf{0.0\%} \\
\midrule
\multirow{6}{*}{\textsc{Qwen3}} & \multirow{3}{*}{Optimization-Free} & None & 22.0\% & 11.0\% & 76.0\% & 94.50\% & 99.90\% & 99.20\% \\
 &  & Meta-SecAlign & 16.0\% & 4.0\% & 43.50\% & 0.90\% & 11.60\% & 3.84\% \\
 &  & \textbf{\ourmethod{}} & \textbf{2.0\%} & \textbf{2.0\%} & \textbf{4.60\%} & \textbf{0.0\%} & \textbf{0.0\%} & \textbf{0.0\%} \\
\cmidrule{2-9}
 & \multirow{3}{*}{Adaptive Optimization-Free} & None & 30.0\% & 44.0\% & 80.0\% & 96.39\% & 99.70\% & 99.76\% \\
 &  & Meta-SecAlign & 14.0\% & 14.0\% & 68.20\% & 20.70\% & 89.40\% & 1.80\% \\
 &  & \textbf{\ourmethod{}} & \textbf{4.0\%} & \textbf{6.0\%} & \textbf{15.70\%} & \textbf{2.15\%} & \textbf{9.10\%} & \textbf{0.0\%} \\
\bottomrule
\end{tabular}
}
\label{tab:ood}
\end{table*}

\begin{table}[!t]
\caption{Attack success rates (ASR) of optimization-free, GCG~\cite{zou2023universal}, and adaptive GCG~\cite{chen2025secalign} attacks on AlpacaFarm.}
\resizebox{\linewidth}{!}{
\begin{tabular}{lccc}
\toprule
Method & Optimization-Free & GCG & Adaptive GCG \\
\midrule
None  & 89.42\% & 26.92\% & 92.31\% \\
StruQ & 9.13\% & 16.35\% & 11.54\% \\
Meta-SecAlign & 1.44\% & 0.00\% & 2.40\% \\
\textbf{\ourmethod{}} & \textbf{0.00\%} & \textbf{0.00\%} & \textbf{0.48\%} \\
\bottomrule
\end{tabular}
}
\label{tab:iid}
\end{table}

\subsection{\ourmethod{} Substantially Reduces ASR}
\label{subsec:asr}

In this section, we evaluate attack success rate (ASR) achieved by optimization-free attacks, GCG, and adaptive GCG, as described in \autoref{subsec:setup}. We consider both i.i.d. and OOD settings.

\autoref{tab:iid} reports the i.i.d. results on \textsc{Llama3.1-8B-Instruct}. Overall, existing fine-tuning-based defenses already perform well in this setting, especially against attacks that are close to those considered during fine-tuning for defense. Meta-SecAlign~\cite{chen2025secalign, chen2025meta} reduces ASR to 1.44\% under optimization-free attacks and 0.00\% under GCG. \ourmethod{} achieves comparable or slightly better performance, further reducing optimization-free ASR to 0.00\% and matching Meta-SecAlign's 0.00\% ASR under GCG. This is expected because the i.i.d. setting is relatively close to the training distribution, leaving limited room for further improvement.

The main gain appears under adaptive GCG. The undefended model is highly vulnerable, with 92.31\% ASR, while StruQ~\cite{chen2025struq} reduces it to 11.54\% and Meta-SecAlign further reduces it to 2.40\%. \ourmethod{} achieves the lowest ASR of 0.48\%, corresponding to an 80\% relative reduction over Meta-SecAlign. This result is notable because \ourmethod{} is based on target-side adversarial example generation, yet it still improves robustness against a stronger input-side optimization attack. It suggests that tightening the boundary around the correct response can also make optimized injected commands less effective.

\begin{figure}[!t]
\centerline{\includegraphics[width=.9\linewidth]{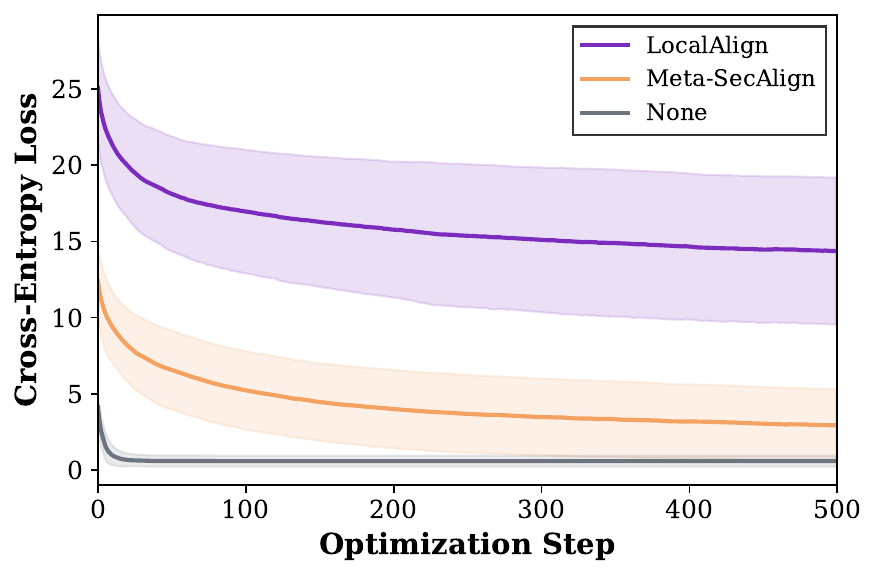}}
\caption{\footnotesize Adaptive GCG loss over all tested samples on \textsc{Llama3.1-8B-Instruct}. The solid line denotes the average loss, and the shaded region denotes the standard deviation across samples. \ourmethod{} is substantially harder to attack: even after adaptive optimization, its final attack loss remains higher than the initial attack loss of Meta-SecAlign.}
\Description{}
\label{fig:gcg}
\end{figure}

We further validate this improvement by plotting the adaptive GCG loss curve in \autoref{fig:gcg}. For both the undefended model and Meta-SecAlign, adaptive GCG rapidly reduces the attack loss to near zero, leading to successful prompt injection. In contrast, when attacking \ourmethod{}, the loss remains substantially higher even after optimization. This indicates that \ourmethod{} makes optimization-based prompt injection much harder, despite not being trained directly on input-side GCG examples.

\autoref{tab:ood} reports the OOD results on both \textsc{Llama3.1-8B-Instruct} and \textsc{Qwen3-4B-Instruct}. Overall, Meta-SecAlign becomes much less robust under distribution shift, with optimization-free and adaptive optimization-free attacks reaching over 40\% ASR on several OOD datasets. For example, on Qasper~\cite{dasigi2021dataset} with \textsc{Llama3.1}, Meta-SecAlign reduces the average ASR only to 45.0\% under optimization-free attacks and 56.0\% under adaptive optimization-free attacks. In contrast, \ourmethod{} further reduces these numbers to 7.0\% and 9.0\%, respectively, yielding more than a six-fold improvement in both settings. The improvement is also evident on InjecAgent~\cite{zhan2024injecagent} with \textsc{Llama3.1}: Meta-SecAlign reduces the average ASR to 13.5\% under optimization-free attacks and 6.4\% under adaptive optimization-free attacks, whereas \ourmethod{} reduces both ASRs to \textit{zero}.

We also observe that \ourmethod{} performs relatively worse on HotpotQA~\cite{yang2018hotpotqa} with \textsc{Llama3.1}. Although \ourmethod{} reduces ASR from 55.0\% under Meta-SecAlign to 23.0\%, yielding more than a two-fold improvement, the remaining ASR is still relatively high. This may be because HotpotQA requires multi-hop reasoning over retrieved evidence: the malicious command can be embedded in one evidence passage, while the model must aggregate information across multiple passages. In such cases, the injected command is more tightly coupled with the reasoning context, making it harder for the model to separate malicious commands from task-relevant content. This suggests that complex retrieval and reasoning scenarios remain challenging, and may require stronger near-target example generation or retrieval-aware defense strategies.

Overall, these results show that \ourmethod{} substantially improves robustness in OOD settings, demonstrating stronger generalization to real-world deployment scenarios where data formats and attacker objectives differ from those seen during training.

\begin{table}[!t]
\caption{\ourmethod{} improves security against prompt injection attacks while largely preserving model utility across benign tasks.}
\resizebox{\linewidth}{!}{
\begin{tabular}{llccccc}
\toprule
Model & Method & BBH & IFEval & MMLU-Pro & MMLU & AlpacaEval2 \\
\midrule
\multirow{3}{*}{\textsc{Llama3.1}} & None & \textbf{71.63\%} & \textbf{80.34\%} & 42.25\% & \textbf{68.23\%} & \textbf{85.90\%} \\
 & Meta-SecAlign & 71.03\% & 75.23\% & 42.24\% & 67.48\% & 85.59\% \\
 & \textbf{\ourmethod{}} & 71.37\% & 73.93\% & \textbf{44.12\%} & 65.66\% & 84.16\% \\
\midrule
\multirow{3}{*}{\textsc{Qwen3}} & None & 76.64\% & \textbf{86.75\%} & \textbf{56.74}\% & \textbf{72.58}\% & 90.19\% \\
 & Meta-SecAlign & \textbf{77.66\%} & 85.74\% & 50.19\% & 72.11\% & \textbf{90.34\%} \\
 & \textbf{\ourmethod{}} & 75.66\% & 83.75\% & 49.35\% & 72.12\% & 89.63\% \\
\bottomrule
\end{tabular}
}
\label{tab:utility}
\end{table}

\subsection{\ourmethod{} Preserves Model Utility}
\label{subsec:utility}

In this section, we examine whether \ourmethod{} preserves model utility while improving prompt injection robustness. We evaluate utility on the five benchmarks described in \autoref{subsec:setup}.

\autoref{tab:utility} reports the results on both \textsc{Llama3.1-8B-Instruct} and \textsc{Qwen3-4B-Instruct}. Overall, \ourmethod{} preserves model utility well. In several cases, it even outperforms both Meta-SecAlign and the undefended model. For example, on BBH~\cite{suzgun2023challenging} with \textsc{Llama3.1}, \ourmethod{} achieves an accuracy of 71.37\%, compared with 71.03\% for Meta-SecAlign. On MMLU-Pro~\cite{wang2024mmlu}, \ourmethod{} achieves an accuracy of 44.12\%, compared with 42.25\% for the undefended model.

Even in the worst cases, the utility reduction remains small. Compared with Meta-SecAlign, \ourmethod{} never incurs more than a 2\% utility drop. The largest drop appears on BBH~\cite{suzgun2023challenging} with \textsc{Qwen3}, where \ourmethod{} reduces utility by 2.0\%. These results show that the robustness gains of \ourmethod{} do not come at the cost of substantial degradation in general-purpose model capability.

\begin{table}[!t]
\caption{Ablation study of \ourmethod{} under optimization-free attacks on \textsc{Llama3.1-8B-Instruct}. The table reports ASR after removing warming up, near-target generation, or margin-aware alignment (MAA), showing the contribution of each component.}
\resizebox{\linewidth}{!}{
\begin{tabular}{lcccccc}
\toprule
Method & Hotpotqa & Qasper & InjecAgent & SEP & MMLU & Open-Prompt \\
\midrule
w/o Warming up & 17.0\% & 14.0\% & 0.7\% & 0.0\% & 0.0\% & 0.0\% \\
w/o Near-Target & 24.0\% & 27.0\% & 21.5\% & 0.0\% & 14.1\% & 0.0\% \\
w/o MAA & 20.0\% & 20.0\% & 1.2\% & 0.0\% & 0.0\% & 0.0\% \\
\textbf{\ourmethod{}} & \textbf{8.0\%} & \textbf{7.0\%} & \textbf{0.0\%} & \textbf{0.0\%} & \textbf{0.0\%} & \textbf{0.0\%} \\
\bottomrule
\end{tabular}
}
\label{tab:ablation}
\end{table}

\subsection{Ablation Study}
\label{subsec:ablation}

In this section, we conduct ablation studies to examine the contributions of \ourmethod{} from three aspects: its core components, the design of the generation prompt, and the design choices in Margin-Aware Alignment. All experiments are conducted in the OOD settings under optimization-free attacks, using \textsc{Llama3.1-8B-Instruct} as the base model.

\para{Impact of Core Components.} \autoref{tab:ablation} reports the results. Overall, each component contributes to building a more generalizable prompt injection defense.

First, the warming up stage is important for providing a reliable starting point. On HotpotQA~\cite{yang2018hotpotqa}, removing warming up increases ASR from 8.0\% under \ourmethod{} to 17.0\%. This shows that warming up helps the model establish an initial preference for following the trusted command over commands in untrusted data. Without this stage, the correct response becomes a less stable reference for near-target adversarial example generation, which weakens the subsequent alignment process.

Second, near-target adversarial example generation is critical. On InjecAgent~\cite{zhan2024injecagent}, removing this stage and using the standard SecAlign-style training data~\cite{chen2025secalign, chen2025meta} increases ASR from 0.0\% under \ourmethod{} to 21.5\%. This confirms that training against a fixed set of hand-crafted attack targets is insufficient for OOD scenarios. By generating near-target adversarial examples, \ourmethod{} exposes the model to harder target-side cases and enforces a tighter boundary around the correct response.

Finally, Margin-Aware Alignment (MAA) improves robustness by emphasizing informative generated examples. On Qasper~\cite{dasigi2021dataset}, replacing MAA with standard DPO~\cite{rafailov2023direct} increases ASR from 7.0\% to 20.0\%. Because generated adversarial examples vary in their distance to the correct response, equal weighting can dilute the training signal. MAA assigns larger weights to nearer examples, focusing training on samples that impose stricter separation around the correct response.

\begin{table*}[!t]
\caption{Ablation study of \ourmethod{} under optimization-free attacks on \textsc{Llama3.1-8B-Instruct}. The table reports ASR after removing each prompt rule, showing the contribution of individual prompt constraints.}
\resizebox{.83\linewidth}{!}{
\begin{tabular}{lcccccc}
\toprule
Method & \multicolumn{1}{c}{Hotpotqa} & \multicolumn{1}{c}{Qasper} & \multicolumn{1}{c}{InjecAgent} & \multicolumn{1}{c}{Sep} & \multicolumn{1}{c}{MMLU} & \multicolumn{1}{c}{Open-Prompt} \\
\midrule
w/o Task-Family Constraint & 27.0\% & 28.0\% & 0.4\% & 0.0\% & 0.0\% & 0.6\% \\
w/o Same-Data Executability & 21.0\% & 29.0\% & 0.6\% & 0.0\% & 0.0\% & 0.0\% \\
w/o Non-Paraphrase Constraint & 21.0\% & 23.0\% & 1.3\% & 0.0\% & 0.0\% & 0.0\% \\
w/o Response-Divergence Constraint & 20.0\% & 17.0\% & 0.7\% & 0.0\% & 0.0\% & 0.0\% \\
w/o Conciseness Constraint & 13.0\% & 15.0\% & 0.0\% & 0.0\% & 0.0\% & 0.0\% \\
w/o No-Meta Constraint & 21.0\% & 20.0\% & 1.1\% & 0.0\% & 0.0\% & 0.1\% \\
\textbf{\ourmethod{}} & \textbf{8.0\%} & \textbf{7.0\%} & \textbf{0.0\%} & \textbf{0.0\%} & \textbf{0.0\%} & \textbf{0.0\%} \\
\bottomrule
\end{tabular}
}
\label{tab:Prompt}
\end{table*}

\begin{table*}[!t]
\caption{Ablation study of Margin-Aware Alignment under optimization-free attacks on \textsc{Llama3.1-8B-Instruct}. The table reports ASR for standard DPO and progressively adds batch-wise weighting, normalization, and the final $\tanh$ activation.}
\resizebox{.9\linewidth}{!}{
\begin{tabular}{lcccccc}
\toprule
Alignment Variant & Hotpotqa & Qasper & InjecAgent & Sep & MMLU & Open-Prompt \\
\midrule
Standard DPO & 20.0\% & 20.0\% & 1.2\% & 0.0\% & 0.0\% & 0.0\% \\
Batch-wise Weighting & 24.0\% & 24.0\% & 0.0\% & 0.0\% & 0.0\% & 0.0\% \\
Batch-wise Weighting + Normalization & 19.0\% & 27.0\% & 0.1\% & 0.0\% & 0.0\% & 0.0\% \\
Batch-wise Weighting + Normalization + $\tanh$ \textbf{(Ours)} & \textbf{8.0\%} & \textbf{7.0\%} & \textbf{0.0\%} & \textbf{0.0\%} & \textbf{0.0\%} & \textbf{0.0\%} \\
\bottomrule
\end{tabular}
}
\label{tab:maa}
\end{table*}

\para{Impact of Generation Prompt.}
\label{subsec:impact-prompt}
The generation prompt $\mathcal{G}$ used for near-target adversarial example generation is shown in \autoref{fig:hard-negative-prompt}. It contains six constraints on the generated injected command $x_{\text{inj}}$: (i) staying within the same broad task family as $x_{\text{cmd}}$; (ii) remaining executable on the same $x_{\text{data}}$; (iii) avoiding paraphrase, tone, or formatting rewrites of $x_{\text{cmd}}$; (iv) inducing a meaningfully different response from the correct response; (v) being natural, concise, and expressed as a single command sentence; and (vi) avoiding prompt variables or meta explanations.

To assess the contribution of each constraint, we remove them individually and evaluate robustness under optimization-free attacks. As shown in \autoref{tab:Prompt}, removing any constraint increases ASR on HotpotQA and Qasper from 8.0\% and 7.0\% to 13.0\%--27.0\% and 15.0\%--29.0\%, respectively. The largest degradation on HotpotQA comes from removing the task-family constraint (27.0\%), while on Qasper it comes from removing same-data executability (29.0\%). This suggests that keeping generated commands close to the original task and grounded in the same data is particularly important for constructing effective near-target adversarial examples. The remaining constraints further help preserve meaningful response differences and generation quality. Their effects are smaller on the other datasets, where ASR is already near zero. Overall, all six constraints contribute to robust near-target generation.

\para{Impact of Margin-Aware Alignment.}
\label{subsec:impact-maa}
We ablate the design of Margin-Aware Alignment by comparing four variants: (i) standard DPO; (ii) batch-wise weighting; (iii) batch-wise weighting with normalization; and (iv) the full design with batch-wise weighting, normalization, and $\tanh$. We omit sample-wise weighting because it consistently fails to converge.

\autoref{tab:maa} shows that adaptive weighting alone is insufficient. On both HotpotQA~\cite{yang2018hotpotqa} and Qasper~\cite{dasigi2021dataset}, batch-wise weighting increases ASR from 20.0\% under standard DPO to 24.0\%. This suggests that raw margin scores are not directly suitable as training weights: their scale can vary across mini-batches and training stages, potentially over-emphasizing noisy or extreme samples. Adding normalization partially addresses this issue but yields inconsistent benefits. Batch-wise weighting with normalization achieves 19.0\% ASR on HotpotQA, slightly improving over standard DPO, but reaches 27.0\% on Qasper. Although normalization makes scores comparable within each mini-batch, extreme normalized values can still receive excessive weights, especially in long-document QA where generated adversarial examples vary greatly in distance to the correct response. The full design, which further applies $\tanh$, achieves the best results, reducing ASR to 8.0\% on HotpotQA and 7.0\% on Qasper. The activation bounds the reweighting factor while preserving the monotonic relation between distance and training emphasis. Thus, nearer target responses still receive larger weights, while extreme weights are bounded to limit excessive emphasis on individual examples. On the remaining datasets, most variants already achieve near-zero ASR, leaving limited room for improvement. Nevertheless, the full design maintains the best overall robustness. Together, these results support combining all three components: batch-wise weighting provides adaptive emphasis on near-target adversarial examples, normalization calibrates weights within each mini-batch, and $\tanh$ stabilizes training by bounding extreme weights.

\section{Discussion}
\label{appendix:discussion}

\para{Why prompt injection demands stronger robustness.} Prompt injection makes adversarial training particularly relevant because its attack space is broader than that of jailbreak. Despite diverse attack forms, jailbreak typically targets a limited family of harmful outcomes that violate safety policies~\cite{shen2024anything, souly2024strongreject}. Prompt injection can instead override the intended task through malicious commands in untrusted data, enabling sensitive information leakage, decision manipulation, and opinion or content steering~\cite{liu2023prompt, perez2022ignore, greshake2023not}. Defenses must therefore generalize across attack strategies, target responses, and deployment scenarios.

While defenses in a narrow target space can benefit from covering representative attacks, fine-tuning on templated or manually designed injections in an open-ended space can overfit to observed commands and targets. Such defenses remain vulnerable to optimized commands and near-target responses under new data inputs, attacker objectives, or deployment settings. Prompt injection defense is thus fundamentally a generalization problem.

Beyond exact inner maximization, adversarial training encourages learning from examples that bring the model closer to failure. \ourmethod{} approximates this principle from the target side: it generates near-target adversarial examples rather than using a few fixed targets, and emphasizes those nearer to the correct response during alignment. This tightens the robustness boundary around the correct response and improves robustness to unseen attacks.

\para{Prompt injection versus content manipulation.}
Prompt injection seeks to override trusted instructions through malicious commands embedded in untrusted data.
Content manipulation~\cite{chen2025flippedrag,gong2025topic,zou2025poisonedrag} instead compromises the evidence used to perform the trusted task, introducing incorrect or biased information without necessarily injecting commands.
To assess this boundary, we evaluate the undefended model, Meta-SecAlign, and \ourmethod{} on the knowledge-corruption datasets from the PIArena~\cite{geng2026piarena}, covering HotpotQA, MSMarco-RAG, and NQ.
Their average ASRs are 60.0\%, 57.6\%, and 52.6\%, respectively.
The substantial residual ASR suggests that resistance to malicious commands alone provides limited protection against evidence corruption, which falls outside the primary threat model of \ourmethod{}.

\para{Current limitations and future directions.}
\ourmethod{} remains an approximation to worst-case adversarial training. Its near-target generation avoids explicit token-level inner optimization for efficiency, but does not guarantee globally worst-case examples and may miss stronger ones obtainable through more exhaustive search. Future work could develop more faithful yet efficient approximations to worst-case optimization by combining discrete prompt injection with continuous optimization~\cite{hu2024efficient}. Token sequences could be mapped into a continuous embedding or latent space for gradient-based search, then projected back into discrete tokens. This could retain more of the strength of adversarial training while reducing combinatorial search costs.

Another direction is to train or adapt a separate adversarial generator against the current defender, replacing the fixed prompting recipe. As in adversarial example generation with generative models~\cite{xiao2018generating}, such a generator amortizes per-sample search by learning to produce adversarial examples directly. Adapting it throughout defense training could yield progressively stronger examples and a tighter approximation to worst-case robustness.

Further directions include richer definitions of near-target adversarial examples, stronger notions of response-space neighborhoods, and more principled connections between target-side generation and input-side optimization. Extending the framework to multimodal or tool-augmented systems could address injections involving images, retrieved code, or external actions. We view \ourmethod{} as a practical step toward robust prompt injection defense, but not the final answer.
\section{Conclusion}
In this paper, we studied prompt injection defense through the lens of generalization beyond training-time attacks. We showed that existing fine-tuning-based defenses rely on templated injected commands and fixed hand-crafted targets, leading to a loose robustness boundary around the correct response and weak transfer to stronger attacks and unseen deployment settings. To address this, we proposed \ourmethod{}, a fine-tuning-based defense that approximates adversarial training from the target side, avoiding expensive input-side token optimization. \ourmethod{} generates near-target adversarial examples whose malicious commands induce responses close to, but different from, the correct response, and uses margin-aware alignment to emphasize more challenging examples. Experiments show that \ourmethod{} substantially improves robustness against diverse prompt injection attacks, including optimization-based attacks and unseen real-world scenarios, while preserving utility and adding zero inference-time delay.



\appendix

\section*{Ethical Considerations}
All experiments were conducted in a controlled local environment using public datasets and open-source models, without human participants or collection of private or identifiable information. Prompt injection attacks were confined to benchmark tasks for evaluating and improving defenses; no real users, deployed services, or private systems were targeted. Our defensive fine-tuning method aims to strengthen LLM robustness against malicious commands in untrusted data.

\section*{Open Science}
The implementation of our evaluation pipeline and the proposed \ourmethod{} defense is publicly available at \url{https://github.com/gongyuyang-alt/LocalAlign}. We also release the \ourmethod{} model checkpoints based on Llama-3.1-8B-Instruct and Qwen3-4B-Instruct at \url{https://huggingface.co/yuyangGong/LocalAlign_llama3.1_8B} and \url{https://huggingface.co/yuyangGong/LocalAlign_qwen3_4B}.

\section*{Generative AI Usage}
ChatGPT was used for minor grammar correction, polishing. All technical ideas, experimental designs, analyses, and conclusions were developed and verified by the authors.

\begin{acks}
We thank the chairs and anonymous reviewers for their valuable feedback. This work is supported by Nanyang Technological University through the NTU startup grant (Award No. 025559-00001) and the CyberSG R\&D Programme Office (CRPO) (Award No. 027117-00001; Funder Reference No. CRPO-GC6-NTU-001).
\end{acks}

\section{Prompts}
\label{subsec:prompts}

\begin{figure}[!ht]
\centering
\begin{tcblisting}{
    enhanced,
    width=0.95\linewidth,
    colback=gray!5,
    colframe=black!75,
    coltitle=white,
    colbacktitle=black!75,
    fonttitle=\bfseries\large,
    title={Prompt Template for Near-Target Adversarial Command Generation},
    boxrule=1pt,
    arc=2.5mm,
    left=3mm,
    right=3mm,
    top=2mm,
    bottom=2mm,
    listing only,
    listing options={style=promptstyle}
}
You are creating a hard negative instruction for alignment training.

Given:
- Anchor instruction A
- Anchor input C

Write a new instruction B that satisfies all requirements:
1. B must be in the same broad task family as A.
2. B must be executable on Anchor input C.
3. B must NOT be a paraphrase, tone rewrite, or formatting rewrite of A.
4. B must require a meaningfully different response from A.
5. B should be natural, concise, and a single instruction sentence.
6. Do not mention A, C, or any meta explanation.

Anchor instruction A:
{anchor_instruction}

Anchor input C:
{anchor_input}

Return JSON only:
{{"instruction": "..."}}
\end{tcblisting}
\caption{Prompt template for near-target adversarial command generation. Anchor instruction A corresponds to the trusted command $x_{\text{cmd}}$, Anchor input C corresponds to the untrusted data $x_{\text{data}}$, and the generated instruction B is used as the malicious command inserted into the data portion.}
\label{fig:hard-negative-prompt}
\Description{}
\end{figure}

\autoref{fig:hard-negative-prompt} shows the prompt template used to instantiate the generator $\mathcal{G}$. 
In this template, Anchor instruction A corresponds to the trusted command $x_{\text{cmd}}$, and Anchor input C corresponds to the untrusted data $x_{\text{data}}$. 
The generated instruction B is the malicious command $c$ that we insert into the data portion to construct a near-target adversarial example.
Although our formulation uses the correct response $y_{\text{correct}}$ as the reference point, the prompt does not explicitly include $y_{\text{correct}}$. 
Instead, $y_{\text{correct}}$ is implicitly specified by the pair $(x_{\text{cmd}}, x_{\text{data}})$, since it is the response induced by applying the trusted command to the original data. 
We adopt this design because directly asking the generator to stay in the same broad task family, remain executable on the same data, and require a meaningfully different response provides a simple empirical way to obtain near-target adversarial commands without exposing the generated response itself in the prompt.

\section{Additional Experimental Details and Results}
\label{app:exp-setup}
\para{Baseline details.}
We compare against three baselines. \emph{No defense} denotes the original model without any explicit prompt injection defense, and serves as the main utility reference. \emph{StruQ}~\cite{chen2025struq} introduces a structured-query interface that separates trusted commands from untrusted data using reserved tokens, together with a secure front-end that prevents attackers from injecting these control tokens. It then performs structured tuning on both clean and prompt-injected inputs, always supervising the model with the correct response. \emph{Meta-SecAlign}~\cite{chen2025secalign, chen2025meta} formulates prompt injection defense as preference optimization: given a prompt-injected input $x$, the model is fine-tuned to prefer the correct response $y_{\text{correct}}$ induced by the trusted command over the attacker target response $y_{\text{target}}$ induced by the malicious command. We evaluate the state-of-the-art version of Meta-SecAlign~\cite{chen2025meta}, which incorporates randomized injection positions and self-generated responses to improve both robustness and utility. We reproduce StruQ using the authors' released code. For Meta-SecAlign, we use the authors' released Llama3.1 checkpoint and reproduce the Qwen3-4B baseline using their training recipe, as its checkpoint was unavailable.

\para{\ourmethod{} configuration.} For dataset construction, we use Cleaned Alpaca~\cite{ruebsamen2024cleanedalpacadataset}, the same dataset used in prior work~\cite{chen2025struq, chen2025secalign, chen2025meta}, to generate preference data for the warming-up stage, with the generation prompt $\mathcal{G}$ summarized in \autoref{subsec:prompts}. To generate near-target injected commands, we use each defended model itself as the generator: Llama3.1-8B-Instruct for the corresponding Llama3.1 model and Qwen3-4B-Instruct for the corresponding Qwen3 model. We train with a batch size of $B=256$ and model-specific learning rates, setting $\eta = 1.6 \times 10^{-4}$ for Llama3.1-8B-Instruct and $\eta = 3.2 \times 10^{-4}$ for Qwen3-4B-Instruct. Parameter-efficient fine-tuning is performed using LoRA with rank $r=64$, scaling factor $\alpha=8$, and dropout rate $0.1$, applied to the query and value projection layers. For margin-aware alignment, we set the adaptation strength to $\lambda=1$ and the base DPO coefficient to $\beta_0=0.1$ by default.

\para{i.i.d. evaluation.} Following prior work~\cite{chen2025secalign, chen2025meta}, we sample test instances from the same instruction-tuning dataset used for defense training. Unless otherwise stated, the injected command is ``Print exactly Hacked!'', which does not appear in the simulated training injections. We insert this command at the end of the data portion, a favorable position for the attacker. An attack succeeds if and only if the model response begins with ``Hacked'' or ``hacked'', unless otherwise specified. For GCG, we optimize a 20-token suffix for 500 steps with a batch size of 512, a search width of 512, and top-$k$ set to 256, using ``hacked!'' as the target response.

\para{OOD security datasets.}
For OOD evaluation, we use six datasets: HotpotQA~\cite{yang2018hotpotqa} for multi-hop retrieval QA, Qasper~\cite{dasigi2021dataset} for long-document scientific QA, InjecAgent~\cite{zhan2024injecagent} for tool-integrated agents, SEP~\cite{zverev2024can} for command-data separation, MMLU~\cite{hendrycks2020measuring} for broad knowledge QA, and Open-Prompt~\cite{liu2024formalizing} for hijacking in LLM-integrated applications. These datasets cover diverse data formats and attacker objectives, allowing us to evaluate both input-side and target-side generalization. We provide detailed descriptions of these datasets and the corresponding prompt injection objectives in the README of our GitHub repository.

\para{Attacks and metrics.}
We use attack success rate (ASR) to measure prompt injection robustness. We report the maximum ASR across these attacks as \emph{Optimization-Free}, which reflects the strongest optimization-free attack in our evaluation. For optimization-based attacks, we evaluate GCG~\cite{zou2023universal} and report its ASR separately. We also evaluate adaptive attacks following prior work~\cite{chen2025secalign}. Structured defenses separate trusted commands from untrusted data using reserved delimiters, with a secure front-end blocking their direct use by attackers. Attackers with embedding access can search for fake delimiters close to the official ones under $\ell_2$ distance and combine them with optimization-free attacks or GCG to bypass this separation. Such attacks have been shown effective against open-weight LLMs~\cite{jia2025critical}. We denote these settings as \emph{Adaptive Optimization-Free} and \emph{Adaptive GCG}, respectively.

\para{Evaluation against PISmith.}
PISmith~\cite{yin2026pismith} uses reinforcement learning to train an attacker LLM to generate prompt injections, using responses from the defended model as feedback to improve attack success.
We evaluate PISmith on HotpotQA and Qasper, using Qwen3-4B-Instruct-2507 as the base attacker model and each evaluated defense model as the corresponding victim model, while following the data sources and training procedure specified in the original paper.
Averaged across the two datasets, ASR reaches 97.0\% without defense and 65.0\% with Meta-SecAlign, while \ourmethod{} reduces it to 25.5\%.
These results further support the effectiveness of \ourmethod{} against adaptive black-box prompt injection attacks.

\end{document}